\documentclass[preprint,12pt]{elsarticle}

\usepackage{multicol}
\usepackage{graphicx}
\usepackage{booktabs}
\usepackage{amssymb,bm,mathrsfs,bbm,amscd}
\usepackage[tbtags]{amsmath}
\usepackage{lastpage}
\usepackage{verbatim}
\usepackage{comment}

\journal{Nuclear Physics A}

\begin{document}

\begin{frontmatter}



\title{Antineutrino flux and spectrum calculation for spent nuclear fuel for the Daya Bay antineutrino experiment \tnoteref{label1}}
 \tnotetext[label1]{Corresponding author}
\author{X.B.Ma\corref{cor1}\fnref{label2}}
 \ead{maxb@ncepu.edu.cn}
\author[label2]{Y.F.Zhao}
\author[label2]{Y.X.Chen}
\author[label3]{W.L.Zhong}
\author[label4]{F.P.An}
\address[label2]{North China Electric Power University, Beijing 102206, China \fnref{label2}}
\address[label3]{Institute of High Energy Physics, Chinese Academy of Sciences, Beijing 100049, China \fnref{label3}}
\address[label4]{East China University of Science and Technology, Shanghai, China \fnref{label3}}
\begin{abstract}
The antineutrino flux from spent nuclear fuel (SNF) is an important source of uncertainty when making estimates of a reactor neutrino flux. However, to determine the contribution from SNF, sufficient data is needed such as the amount of spent fuel in the pool, the time after discharged from the reactor core, the burnup of each assembly, and the antineutrino spectrum of each isotope in the SNF. A method to calculate this contribution is proposed. A reactor simulation code verified against experimental data has been used to simulate fuel depletion by taking into account more than 2000 isotopes and fission products, the quantity of SNF in each of the six spent fuel pools, and the time variation of the antineutrino spectra after SNF discharging from the core. Results show that the SNF contribution to the total antineutrino flux is about 0.26\%--0.34\%, and the shutdown impact is about 20\%. The SNF spectrum alters the softer part of the antineutrino spectra, and the maximum contribution from the SNF is about 3.0\%. Nevertheless, there is an 18\% difference between the line evaluate method and under evaluate method. In addition, non-equilibrium effects are also discussed, and the results are compatible considering the uncertainties.
\end{abstract}


\end{frontmatter}

\section{Introduction}
Reactor antineutrinos are used in the study of neutrino oscillations and in the search for signatures of nonstandard neutrino interactions, as well as to monitor reactor conditions to safeguard operations. Antineutrino flux is an important source of uncertainty associated with measurements in reactor neutrino experiments. The time dependence of the antineutrino flux and spectrum of $\bar{\nu}_{e}$ from a reactor can be estimated using
\begin{equation}
\frac{d^{2}N(E,t)}{dEdt}=\sum_{i}\frac{W_{th}(t)}{\sum_{j}f_{j}(t)e_{j}}f_{i}(t)S_{i}(E)c_{i}^{ne}(E,t)+S_{SNF}(E,t),
\label{flux_equation}
\end{equation}
where $W_{th}(t)$ is the reactor thermal power, $f_{i}(t)$ the fission fraction associated with each isotope, $e_{i}$ the thermal energy release per fission for each isotope, $S_{i}(E)$ a function of the $\bar{\nu}_{e}$ energy $E$ signifying the $\bar{\nu}_{e}$ yield per fission for each isotope, $c_{i}^{ne}(E,t)$ the non-equilibrium correction of the long-live fission fragment isotopes, and $S_{SNF}(E,t)$ the yield from the spent nuclear fuel (SNF). As a result, the antineutrino flux from the SNF and spectra from all its isotopes must be estimated more accurately to provide a precise estimate of the total antineutrino flux and energy spectrum. Antineutrinos emitted from the SNF contribute to the soft part of the energy spectrum\citep{VKopeikin_2004} and introduce a non-negligible systematic uncertainty. Hence study was performed on the impact of the SNF on $\theta_{13}$ sensitivity for the Daya Bay antineutrino experiment\cite{Anfeng_spt}.

In the range of 1.8--4.0~MeV, the contribution of the rate of antineutrino events from SNF was assessed\citep{zhoubin} to be above 4\%. This event rate differed largely from previous results\citep{Anfeng_spt} of below $0.2\%$.

However, the purpose of this study was to calculate the SNF rate and spectrum more precisely. Determining the contribution of SNF precisely is difficult because much data including the number of assemblies in each of the SNF pools, the amount of burnup in each assembly, the rate of discharge from each core, the antineutrino spectrum of each isotope, and accurate simulation data for each assembly, are needed. Previously, a simple model\cite{zhoubin} was used to simulate fuel depletion in the reactor but did not consider the variation in neutron flux during reactor operation.

A method to calculate the SNF antineutrino rate and spectrum is proposed. In this method, a reactor simulation code verified using experimental data is used to simulate fuel depletion by taking into account more than 2000 isotopes and fission products, the amount of SNF in each of the storage pools, and the time variation of the antineutrino spectrum for SNF after each core discharge. The SNF rate is about 0.3\% and the shutdown impact is found to be about 20\%.

\section{Neutrino spectrum after shutdown for one batch}
\label{neutrino_spectrum}
The Daya Bay reactor is a pressurized water reactor (PWR) that uses 175 fuel assemblies in the core. At the end of an equilibrium cycle, about one third of the fuel is removed from the reactor core and sent to one of six SNF storage pools. The antineutrino spectrum of the SNF is calculated using
\begin{equation}
\label{rho}
\rho^{r}_{m}(E_{\nu},t)=\sum_{i} A_{i}(t)\rho_{i}(E_{\nu}),
\end{equation}
where $\rho(E_{\nu},t)$ is the total normalized antineutrino spectrum of the SNF of reactor number $r$ and batch number $m$  at time t; A$_{i}$(t) and $\rho_{i}$($E_{\nu}$) are the activity and the normalized antineutrino spectrum of isotope $i$.

Long-lived fission products are chosen based on a threshold criterion related to the inverse beta decay and its half-life; the criterion is E$_{d}\geq$1.8 ~MeV and T$_{1/2}\geq$ 10h as the threshold for inverse beta decay reaction is around 1.8~MeV, and isotopes with a short half-life reach equilibrium right after the reactor is turned on or decay rapidly after being taken out of the reactor. The properties of all these long-lived fission products are given in references \citep{zhoubin} and \citep{Anfp_phd}.

\begin{figure}
\begin{center}
\includegraphics[width=10cm]{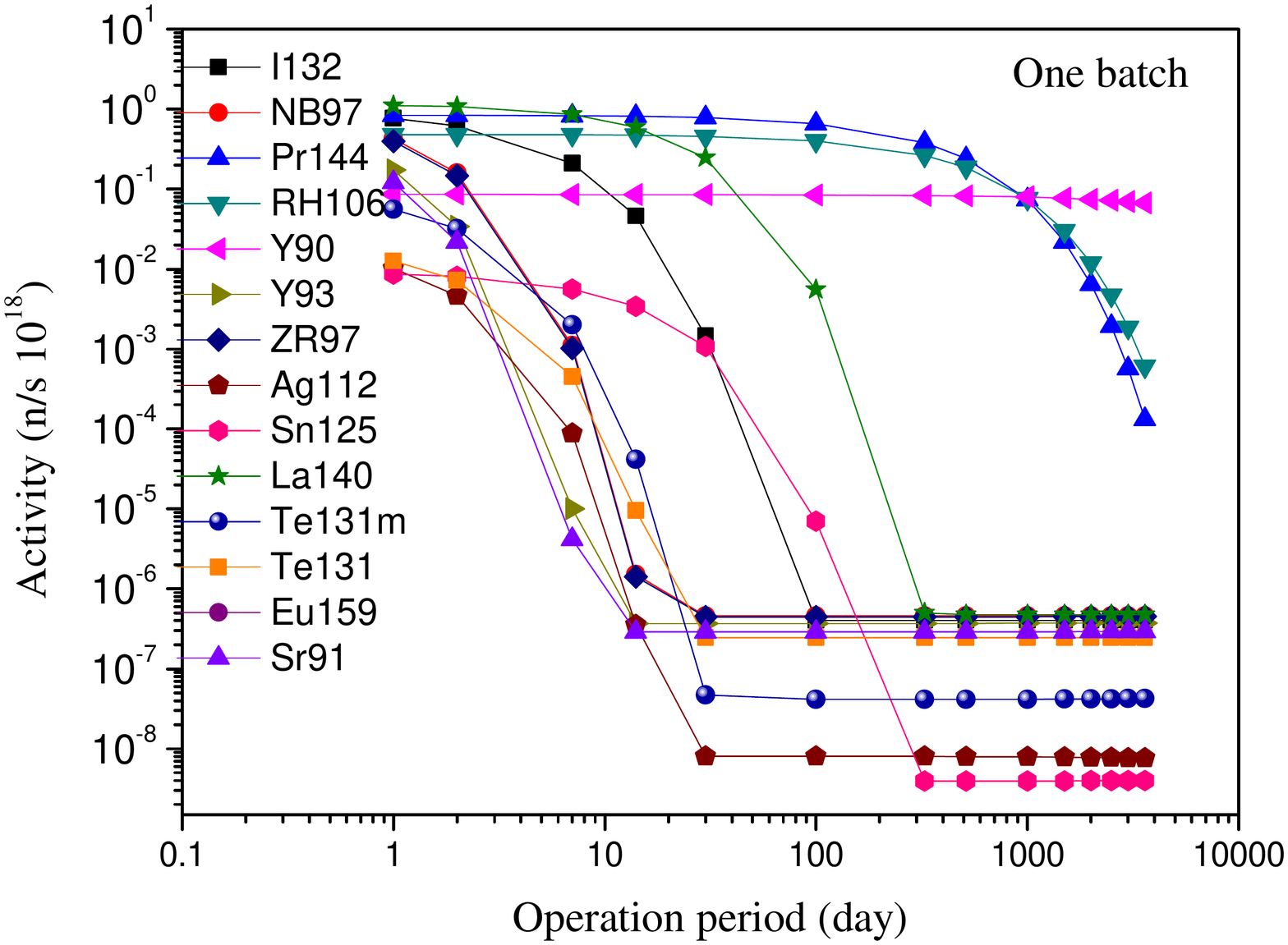}
\caption{Time-dependence after shutdown of the activities in one batch of long-lived fission products}
\label{activity11}
\end{center}
\end{figure}

To obtain A$_{i}$(t), calculations were performed that summed at each instant the contributions from individual fission fragments with allowance for their yields from the fission process, decay diagrams, and life-times. A database containing details of about 571 fission fragments was used\citep{VKopeikin_2001}. For this study, the activity of each long-lived fission product is calculated using ORIGEN-ARP\citep{origenarp} and RMC\citep{rmc} codes by solving the depletion equation. A database containing about 1946 nuclides (including 1119 fission fragments) is used in the depletion calculation, and results from the latest Evaluated Nuclear Data File ENDF/B-VII.0\citep{Endfb7} are used in the transport calculation. The measurements \citep{arprmc1,arprmc2} included 38 SNF samples from fuel irradiated in three different PWRs operating in the United States and Japan. They were used to validate and verify the ORIGEN-ARP, code. The results show that most of the fission products are well predicted within 5$\%$ on average except for $^{150}$Sm and $^{151}Sm$, for which there is a systematic overestimation in the 30$\%$ range. However, the decay energies E$_{d}$ of $^{150}$Sm and $^{151}Sm$ are less than 1.8~MeV. Hence, they are not important for evaluating the SNF antineutrino rate and spectrum. Neutron flux, energy spectrum, and geometry have big influences non-equilibrium\citep{muller}. The neutron flux varies with fuel burnup, and the geometry obtained from the nuclear power plant has been used to ensure results are more accurate. The activities of long-lived fission products for one batch as a function of time after shutdown are presented in Fig.~\ref{activity11}.

To evaluate the antineutrino spectrum of the SNF $\rho(E,t)$, the normalized antineutrino spectrum of all isotopes is calculated. The beta-decay spectrum $\rho(E_{\nu})$ for a single transition in nucleus (Z,A) with end-point energy $E_{0}=E_{e}+E_{\nu}$ is assumed to have the form\citep{hayes}
\begin{equation}
\rho(E_{e})=k(E_{0},Z)E_{e}p_{e}(E_{0}-E_{e})^{2}F(Z+1,E_{e})C(E_{e})[1+\delta(E_{e},Z,A)],
\label{equation-2}
\end{equation}
where $k(E_{0},Z)$ is the normalization constant, $E_{e}$ and $p_{e}$ are the full electron energy and momentum, and $F(Z+1,E_{e})$ describes the Coulomb effect on the outgoing electron with the charge of the daughter nucleus, and $C(E_{e})$ is a shape factor for forbidden transitions due to additional lepton momentum. If the radiative correction is neglected because it is very small, the term $\delta(E_{e},Z,A)$ represents $\delta_{FS}+\delta_{WM}$. The correction of $\delta(E_{e},Z,A)$ in reference \citep{hayes,patric_huber} has been used. The $\bar{\nu_{e}}$ spectrum as a function of the antineutrino energy $E_{\nu}$ is obtained from Eq.~(\ref{equation-2}) by substituting $E_{\nu}=E_{0}-E_{e}$ in the above formula.

From the time averaged new SNF spectrum (Fig.~2), a discrepancy is recorded at energies above 2.75~MeV that may have been caused by using different simulation tools and different nuclear library.

\begin{figure}
\begin{center}
\includegraphics[width=9cm]{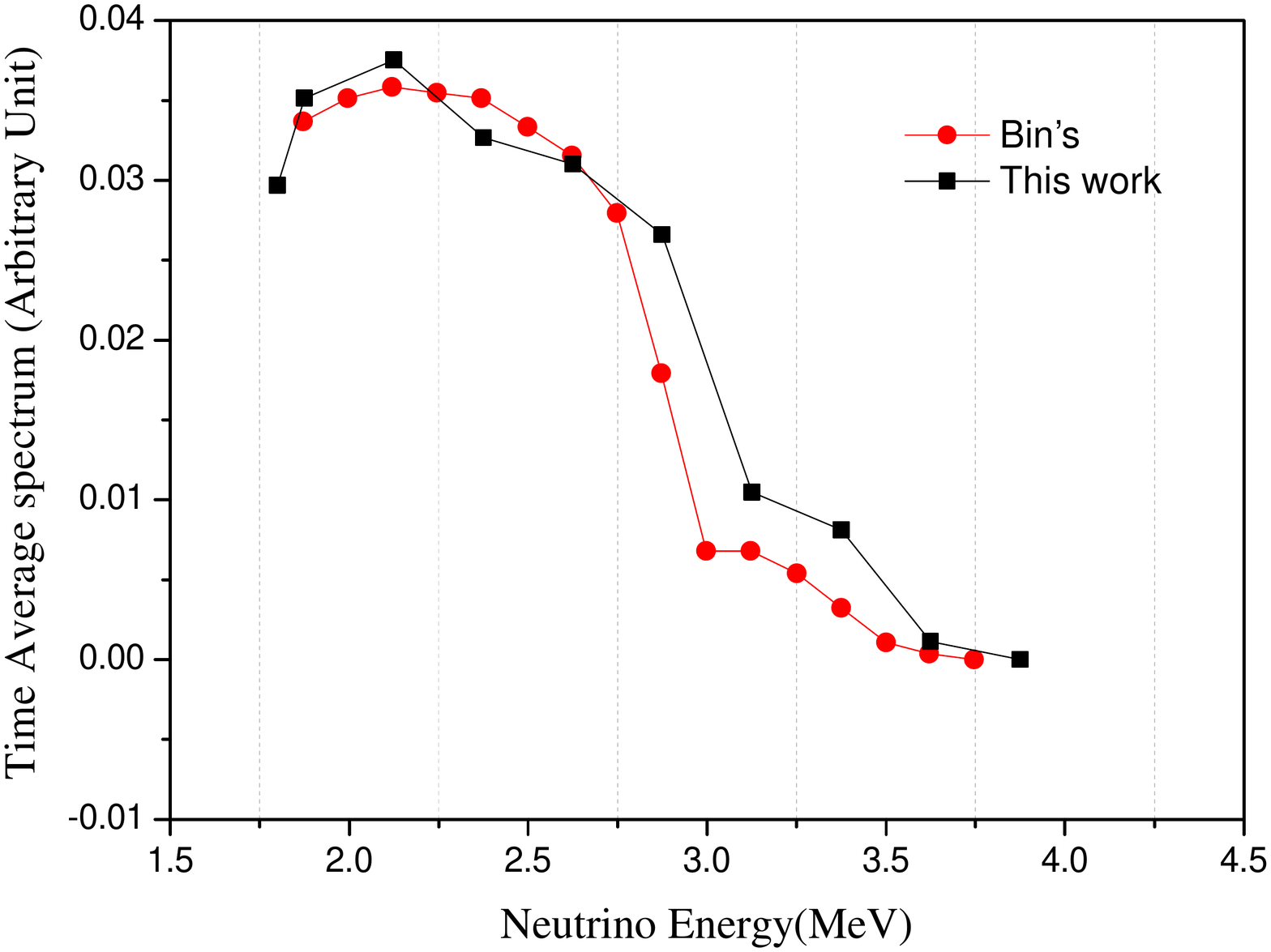}
\caption{Comparison of the new time average SNF spectrum and Zhou Bin's spectrum \citep{zhoubin}}
\label{activity}
\end{center}
\end{figure}
\begin{figure}
\begin{center}
\includegraphics[width=9cm]{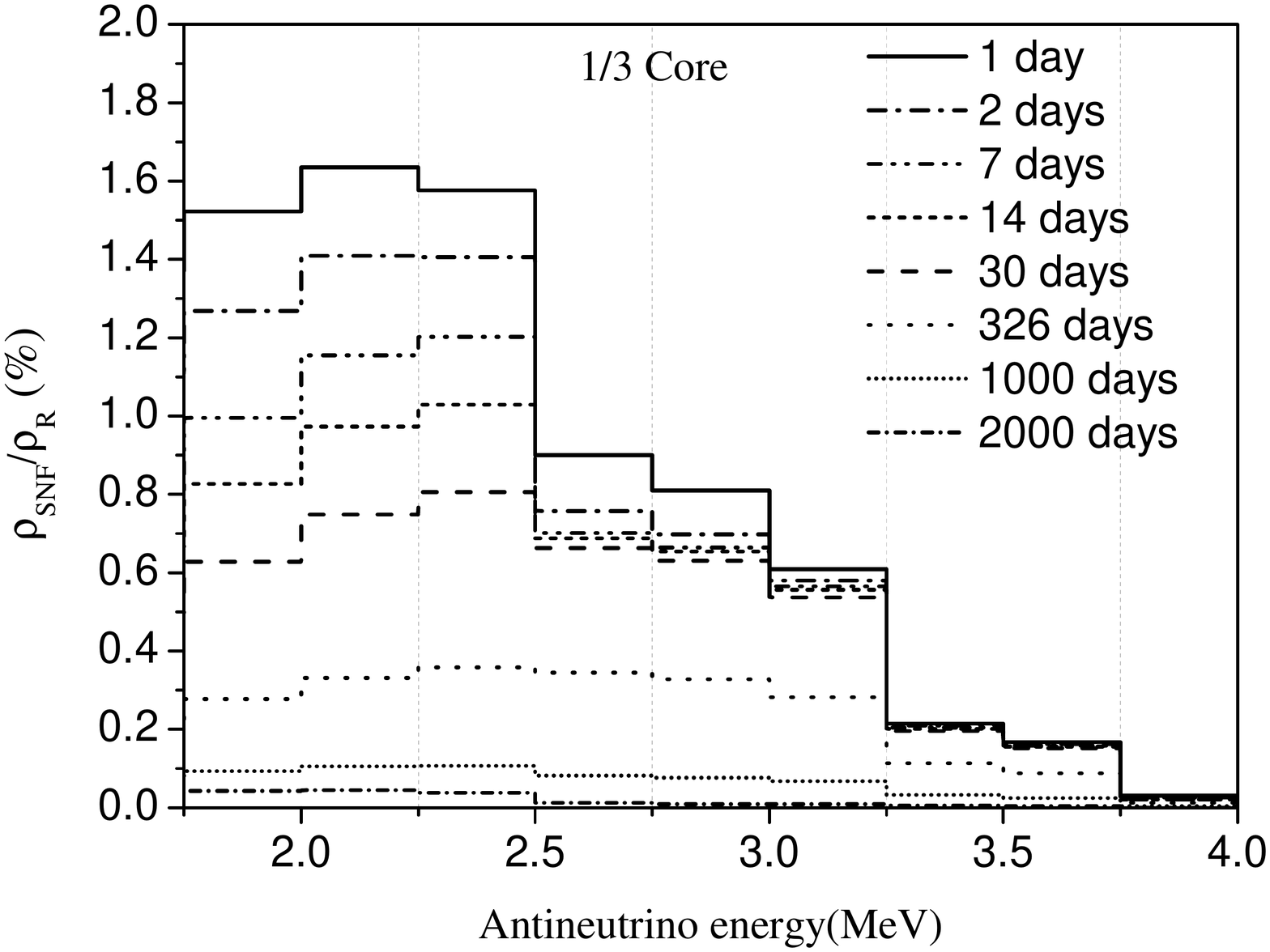}
\caption{$\rho_{SNF}$/$\rho_{R}$ as a function of antineutrino energy}
\label{one_third core}
\end{center}
\end{figure}
The Daya Bay reactor usually employs a $1/3$ refueling strategy with a refueling batch of 52 spent fuel assemblies during each annual refueling cycle. The fuel is spent after three burnup cycles. The antineutrino spectrum of the SNF of one batch (about $1/3$ fuel mass of the whole core) was obtained using Eq.~(\ref{rho}). The ratio of the antineutrino spectrum of the SNF of one batch to that of the whole core under full power operations is shown in Fig.~\ref{one_third core}. The ratio $\rho_{SNF}$/$\rho_{R}$ decreases quickly after the SNF is moved from the reactor core because of activities of some isotopes (short half life constant) decreasing quickly.

The refueling period is usually one to two months, which is usually shorter than the storage period of SNF (usually five years). However, the flux rate in the refueling period is larger and this effect must be taken into account when we calculate the SNF spectrum. The results of shutdown impact are discussed in Section \ref{SNF_section}.

\section{SNF batch ratio}
\label{snf_bath_ratio}
The discharged burnup values for all assemblies are usually different because the burnup histories differ. In general, the larger the burnup is for an assembly, the more antineutrinos are emitted over the same time period because of more fission products. Therefore, the number of antineutrinos emitted from the SNF is assumed to be proportional to the burnup value and the SNF amount. To evaluate the antineutrino flux of the spent fuel, the SNF batch ratio $\alpha(m)$ is defined as
\begin{equation}
\alpha(m)=\sum_{i} Burnup(m,i)/(Burnup_{average}\times N/3),
\label{alpha}
\end{equation}
where Burnup(m,i) is the value of the burnup of assembly $i$ for batch $m$, $Burnup_{average}$ the average burnup of batch $m$, and $N$ the total number of assemblies in the core. The $Burnup_{average}$ is assumed to be equal to 45~GW.d/tU and $N$ is equal to 157. For the LingAo 3 and 4 reactors, the SNF batch ratio is assumed to be 1.0 because detailed burnup data are not known.

\begin{figure}
\begin{center}
\includegraphics[width=9cm]{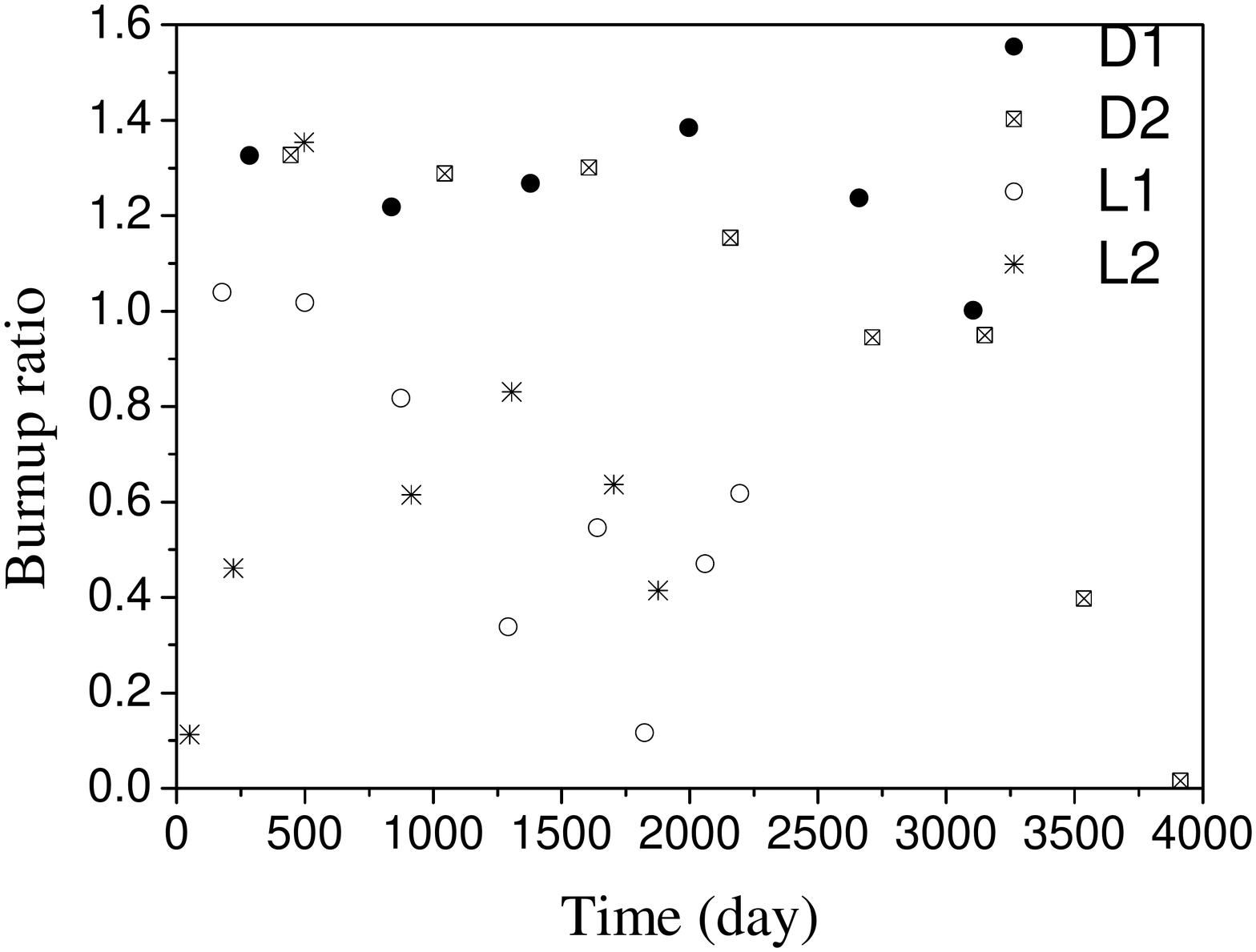}
\caption{Burnup ratio of each reactor vs time after discharge from the core. Dn denotes the Daya Bay n reactor; Ln denotes the LingAo n reactor (n=1,2).}
\label{fig1}
\end{center}
\end{figure}

\section{SNF neutrino spectrum and flux}
\label{SNF_section}
Using equation (\ref{rho}) and (\ref{alpha}), the estimate of the SNF antineutrino flux for a single reactor antineutrino experiment over a given period $[t_{1},t_{2}]$ is
\begin{equation}
\label{equation-x1}
S_{r}^{SNF}(E_{\nu})=\int_{t_{1}}^{t_{2}} \sum_{m}\alpha(m)\rho_{m}^{r}(E_{\nu},t)dt,
\end{equation}
where $r$ is the reactor number, $m$ is the batch number, and $\alpha(m)$ the SNF batch ratio obtained in Section \ref{snf_bath_ratio}, $\rho_{m}^{r}(E_{\nu},t)$ the antineutrino spectrum of $m$ batch and $r$ reactor calculated by Eq.~(\ref{rho}). If the bottom spectrum is used to evaluate the integral, the method called as under evaluate method, and if line interpolation, it is line evaluate
method. Using Eq.~(\ref{equation-x1}), the SNF spectrum at the detector is
\begin{equation}
R_{d}^{SNF}(E_{\nu})= \sum_{r}\frac{1}{4\pi L_{r}^{2}}
[P_{d}(E_{\nu},L_{r}^{d})S_{r}^{SNF}(E_{\nu})\sigma_{IBD}(E_{\nu})]\cdot\epsilon_{d}N_{d},
\label{equation-3}
\end{equation}
where $R_{d}^{SNF}(E_{\nu})$ is the number of detected antineutrino per second with energies between $E_{\nu}$ and $E_{\nu}+dE_{\nu}$, $L_{r}$ the length of the base line, $P_{d}(E_{\nu},L_{r}^{d}))$ the survival probability of an antineutrino with energy $E_{\nu}$ over baseline $L_{r}^{d}$, $\sigma_{IBD}(E_{\nu})$ the cross section for the inverse beta decay reaction, $\epsilon_{d}$ the detector efficiency, and $N_{d}$ the proton number of the detector.

The antineutrino spectrum emitted from a single reactor at a given period is evaluated using equation (\ref{equation-x2}).
\begin{equation}
S_{r}^{Reactor}(E_{\nu})=\int_{t_{1}}^{t_{2}}\sum_{i}f_{i}^{r}(t)S_{i}(E_{\nu})dt,
\label{equation-x2}
\end{equation}
where $S_{i}(E_{\nu})$ is the antineutrino spectrum of isotope i, and $f_{i}^{r}(t)$ the fission fraction of the reactor $r$ for a given isotope $i$ at given time $t$. Using the above equation, the spectrum in the detector can be calculated as
\begin{equation}
R_{d}(E_{\nu})= \sum_{r}\frac{1}{4\pi L_{r}^{2}}\frac{W_{th}^{r}}{\sum_{i}f_{i}^{r}e_{i}}
[P_{d}(E_{\nu},L_{r}^{d})S_{r}^{Reactor}(E_{\nu})\sigma_{IBD}(E_{\nu})]\cdot\epsilon_{d}N_{d},
\label{equation-4}
\end{equation}
where $R_{d}(E_{\nu})$ is the number of antineutrinos per second detected in the detector between energy $E_{\nu}$ and $E_{\nu}+dE_{\nu}$, $W_{th}^{r} (MW)$ is the thermal power of the reactor, and $e_{i}(MeV)$ the fission energy of isotope $i$; the other parameters have the same meaning as in Eq.~(\ref{equation-3}).

\begin{figure}
\begin{center}
\includegraphics[width=10cm]{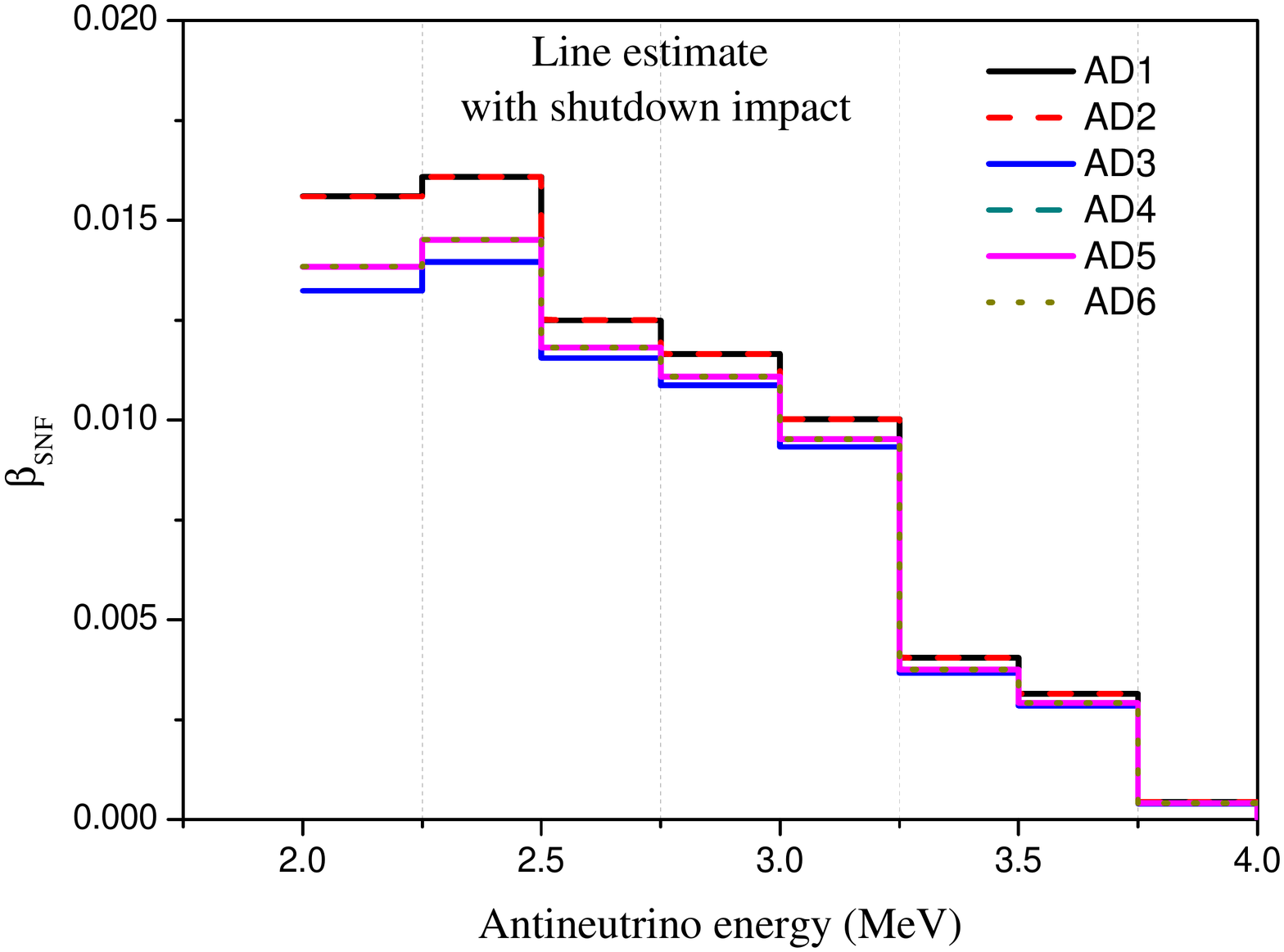}
\caption{ SNF spectrum ratio obtained using the line estimate method for AD1--AD6 over the P12E period (AD is antineutrino detector and the number from 1 to 6 is the detector index)}
\label{line_estimate_shutdown}
\end{center}
\end{figure}

\begin{figure}
\begin{center}
\includegraphics[width=9cm]{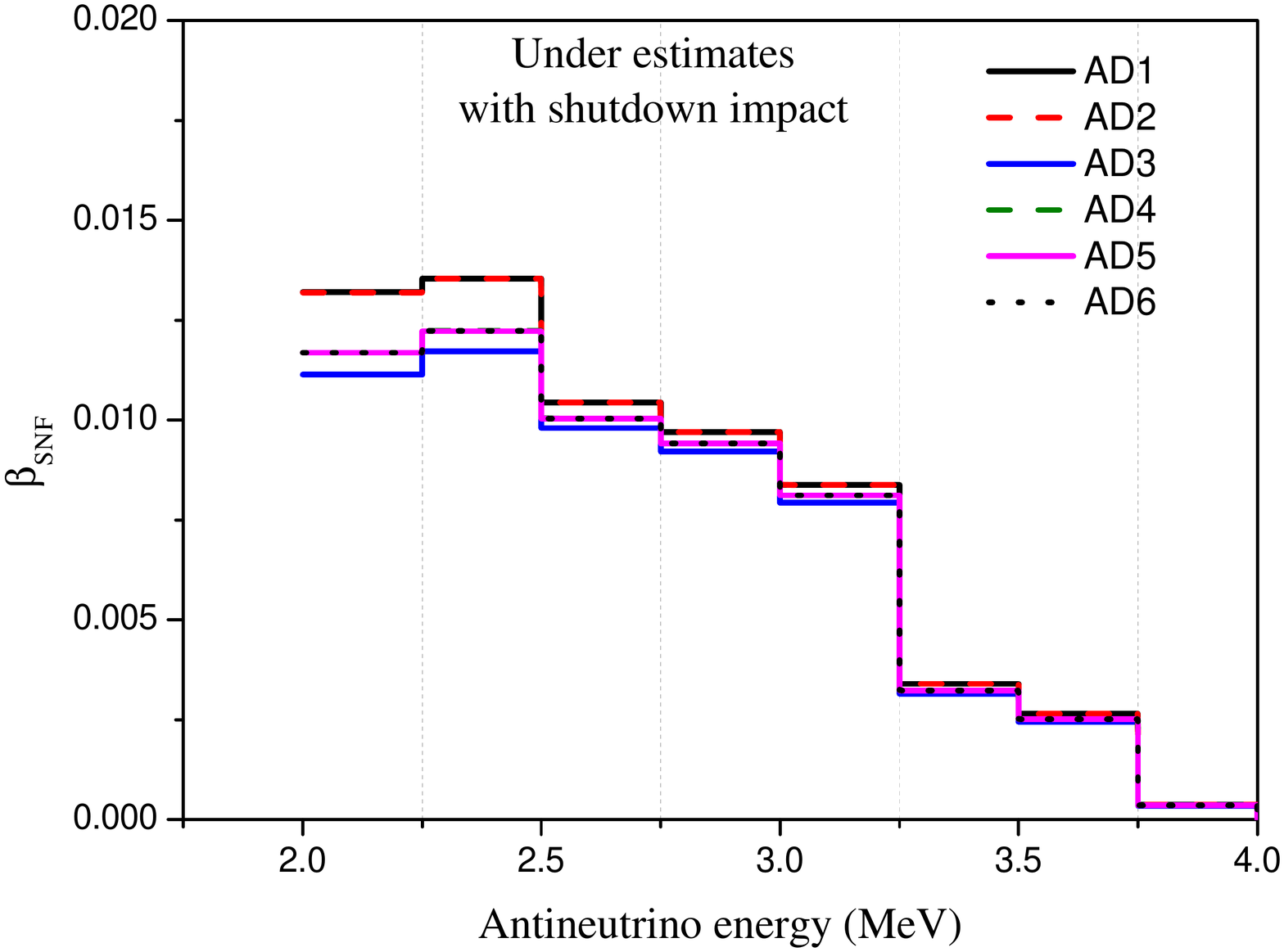}
\caption{SNF spectrum ratio obtained using the under estimate method for AD1--AD6 over P12E period}
\label{under_estimate_shutdown}
\end{center}
\end{figure}

\begin{figure}
\begin{center}
\includegraphics[width=9cm]{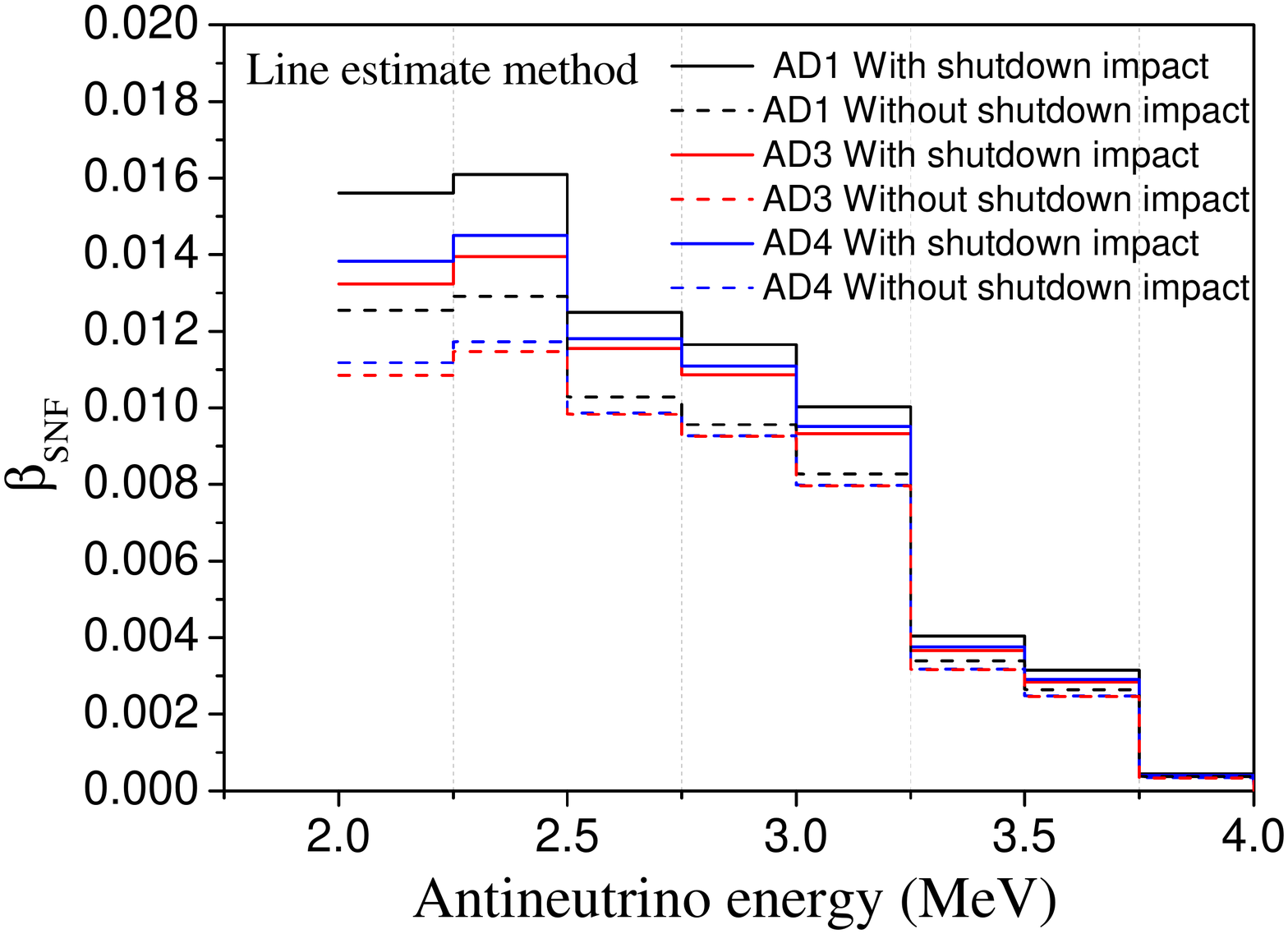}
\caption{Contribution in the antineutrino spectrum when taking shutdown into account}
\label{shutdown_impact}
\end{center}
\end{figure}

The SNF contribution can be represented as a SNF spectrum ratio defined as
\begin{equation}
\beta_{SNF}(E_{\nu}) = \frac{R_{d}^{SNF}(E_{\nu})}{R_{d}^{reactor}(E_{\nu})+R_{d}^{SNF}(E_{\nu})}
\label{SNF_ratio}
\end{equation}
where $R_{d}^{SNF}(E_{\nu})$ and $R_{d}^{reactor}(E_{\nu})$ denote the antineutrino spectra of the SNF and reactor. These spectra have been integrated over P12E period (P12E is the name of Daya Bay neutrino experiment data from 2011/12/24 to 2012/7/28) using the line estimate method and the under estimate method. The former linearly interpolates $\rho_{SNF}$/$\rho_{R}$ between the given time points using Fig.~\ref{one_third core} and the latter uses the lower flux between the given time points. SNF spectrum ratios calculated using both these methods are presented in Figs.~\ref{line_estimate_shutdown} and \ref{under_estimate_shutdown}. The results from the line estimate method are 18\% greater than those from the under estimate method. In general, it takes about one month to reload about $1/3$ new fuel. The contribution of the neutrino which emitted from the other $2/3$ old fuel in the reloading period is called shutdown impact. The impact on shutdown shown in Fig.~\ref{shutdown_impact} indicates about a 20\% increase in the spectrum when taking into account the shutdown effect. These results have been verified in calculations by C. Lewis\citep{CLewis}

\begin{figure}
\begin{center}
\includegraphics[width=9cm]{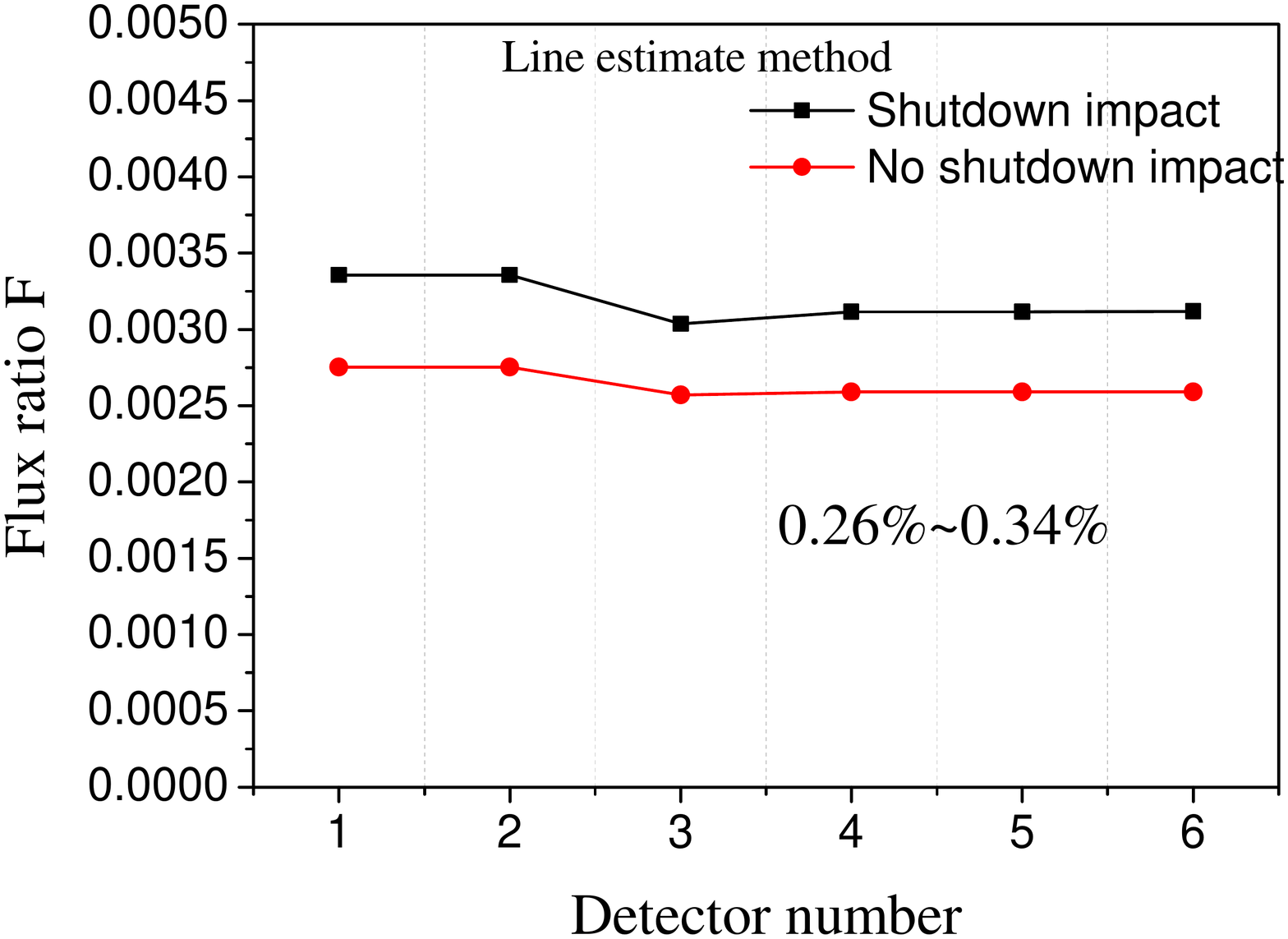}
\caption{Flux ratio obtained from the different detectors}
\label{flux ratio}
\end{center}
\end{figure}

The total contribution of the SNF is quantified by defining the flux ratio
\begin{equation}
F = \frac{\int R_{r}^{SNF}(E_{\nu})dE_{\nu}}{\int (R_{r}^{SNF}(E_{\nu})+R_{r}(E_{\nu}))dE_{\nu}}
= \frac{R_{SNF}}{R_{SNF}+R_{reactor}},
\label{F_equation}
\end{equation}
where $r$ is the reactor index. From a plot of the flux ratio (Fig.~\ref{flux ratio}). The contribution of the flux from the SNF varies between 0.26\% and 0.34\% depending on whether or not shutdown has been taken into account.

From Eq.~(\ref{rho}), the uncertainty associated with the SNF spectrum depends on the uncertainties associated with the isotope activities and the normalized antineutrino spectrum. The uncertainty for each isotope activity arises mainly from the concentration density calculated by the code. The activity uncertainties for the isotopes are 5\% according to the Takahama-3 benchmark calculations\citep{arprmc2}. The normalized antineutrino spectrum can be evaluated to within about 3\% with slight changes at different energies\citep{patric_huber}.

\section{Non-equilibrium corrections}
Spectra from the Institute Laue--Langevin (ILL) reactor were acquired. They were taken after a short irradiation time in conditions when the reactor produced a quasi-pure thermal neutron flux. Depending on the isotopes the measurement period was between 12~h and 1.8~days. However, for the neutrino reactor experiment, the irradiation time period needs to be over a full cycle of the reactor, usually 1.0--1.5~years. Because the irradiation time in the ILL experiment is short, some long-lived fission products cannot reach equilibrium. The neutrinos emitted from these long-lived products should be taken into account in the neutrino flux spectra. Their contribution to the reference neutrino spectra (ILL spectra) is studied and the associated corrections are computed.

The analysis uses Reactor Monte Carlo code (RMC) simulations of a PWR assembly of AFA-3G type exhibiting a moderation ratio equal to that of a PWR core to represent the full core conditions. The infinite multiplication factor has been successfully compared with similar simulations performed with DRAGON . This simulation reproduces very well real physical conditions of a reactor core. Equation~(\ref{rho}) was used to calculate the non-equilibrium antineutrino spectrum. The method used to calculate the non-equilibrium antineutrino spectrum is the same as that for the SNF spectrum except for the RMC simulation time period. In the SNF case, the simulation time covers a long period, which begins when new fuel is loaded in the reactor core to when spent fuel is removed from the core and stored in the pool. However, in the non-equilibrium case, the simulation time covers a burnup cycle. For some relevant low energy bins,the relative non-equilibrium corrections at various times to be applied to the ILL spectrum are listed in Table \ref{tabu235}. As expected, the accumulation of long-lived isotopes accounts for a positive deviation, the amplitude of which decreases with neutrino energy and becomes negligible above 3.5~MeV. Indeed, the non-equilibrium corrections depend on the system parameters (e.g., neutron flux and energy spectrum, geometry, fuel enrichment) used in the calculation. The total error is about 30\% of the total off-equilibrium corrections\citep{muller}. Taking into account these uncertainties, our results are consistent with Mueller's results.
Note that Mueller's reference spectra are calculated using their code with independent fission yields and fixed neutron flux 3 $\times$ 10$^{14}$ $neutron/(cm^{2}.s)$ whereas our reference spectra are those calculated in the RMC code. The fission product with criteria half time $\geqslant$ 1.8MeV and end point energy $\geqslant$10 h were different with Mueller's complex criteria to select the fission product to calculated the non-equilibrium contribution. The difference may be due to the above two reasons.

Non-equilibrium effects were also evaluated in Ref\citep{kopei20041} where the $\beta$ branches of 571 fission fragments were used and fission yields were taken from \citep{kopei20042}. Our results are compatible considering the quoted uncertainties and the possible discrepancies in the neutron energy spectrum and flux used in the calculation. Some additional discrepancies may arise from the smaller number of fission products used in the calculation of Kopeikin et al.\citep{kopei20042}

\begin{table}
\begin{center}
\caption{\label{tabu235}
Relative non-equilibrium correction (in \%) to be applied to the ILL spectrum}
\footnotesize
\begin{tabular*}{100mm}{c@{\extracolsep{\fill}}ccccc}
\toprule
Energy(MeV) & 10 days & 320 days & 452 days & 558.3 days \\ \hline
&&$^{235}$U && \\ \hline
 1.5 & 3.86& 4.14 & 4.20 & 4.25 \\
 1.75 &3.85 &4.20 &4.27& 4.33 \\
 2& 3.55& 3.97& 4.06& 4.14\\
2.25& 2.78&	3.25& 3.36& 3.44\\
2.5& 2.63& 3.09& 3.20& 3.27\\
2.75&2.32& 2.77& 2.88& 2.95\\
3& 1.39& 1.79& 1.90 & 1.98 \\
3.25& 0.20& 0.33&0.38& 0.42 \\
3.5& 0.16 & 0.25& 0.30 &0.33 \\
3.75& 0.020 & 0.035 & 0.041& 0.045\\
4& 0.0& 0.0&0.0& 0.0\\ \hline
&&$^{239}$Pu && \\ \hline
1.5 &1.61& 1.98& 2.04& 2.11 \\
1.75 & 1.33& 1.76& 1.85 & 1.92 \\
2.0 & 0.99 & 1.50& 1.61& 1.70 \\
2.25& 0.54 & 1.11& 1.25 & 1.34\\
2.5& 0.45 & 1.03& 1.18& 1.27 \\
2.75 &0.37 & 0.93& 1.07& 1.16 \\
3.0 & 0.18 & 0.70& 0.85& 0.94 \\
3.25 & 0.020& 0.18&0.25& 0.31 \\
3.5& 0.016& 0.15& 0.20 & 0.24 \\
3.75& 0.002& 0.021& 0.029& 0.035 \\
4& 0.0& 0.0&0.0& 0.0\\ \hline
&&$^{241}$Pu && \\ \hline
1.5 & 1.61& 2.00& 2.04& 2.11\\
1.75& 1.33 & 1.76 & 1.85 & 1.92\\
2&0.98 & 1.50& 1.61 & 1.70\\
2.25& 0.54 & 1.11& 1.25 &1.34\\
2.5&0.45 &1.03&1.18 & 1.268\\
2.75 & 0.37& 0.93& 1.07& 1.17 \\
3 & 0.18 & 0.70 & 0.85& 0.94\\
3.25& 0.02 & 0.18 & 0.25& 0.31 \\
3.5 & 0.016 & 0.15 & 0.21& 0.24\\
3.75 & 0.002& 0.021& 0.029& 0.035\\
4& 0.0& 0.0&0.0& 0.0\\
\bottomrule
\end{tabular*}
\end{center}
\end{table}

\begin{figure}
\begin{center}
\includegraphics[width=9cm]{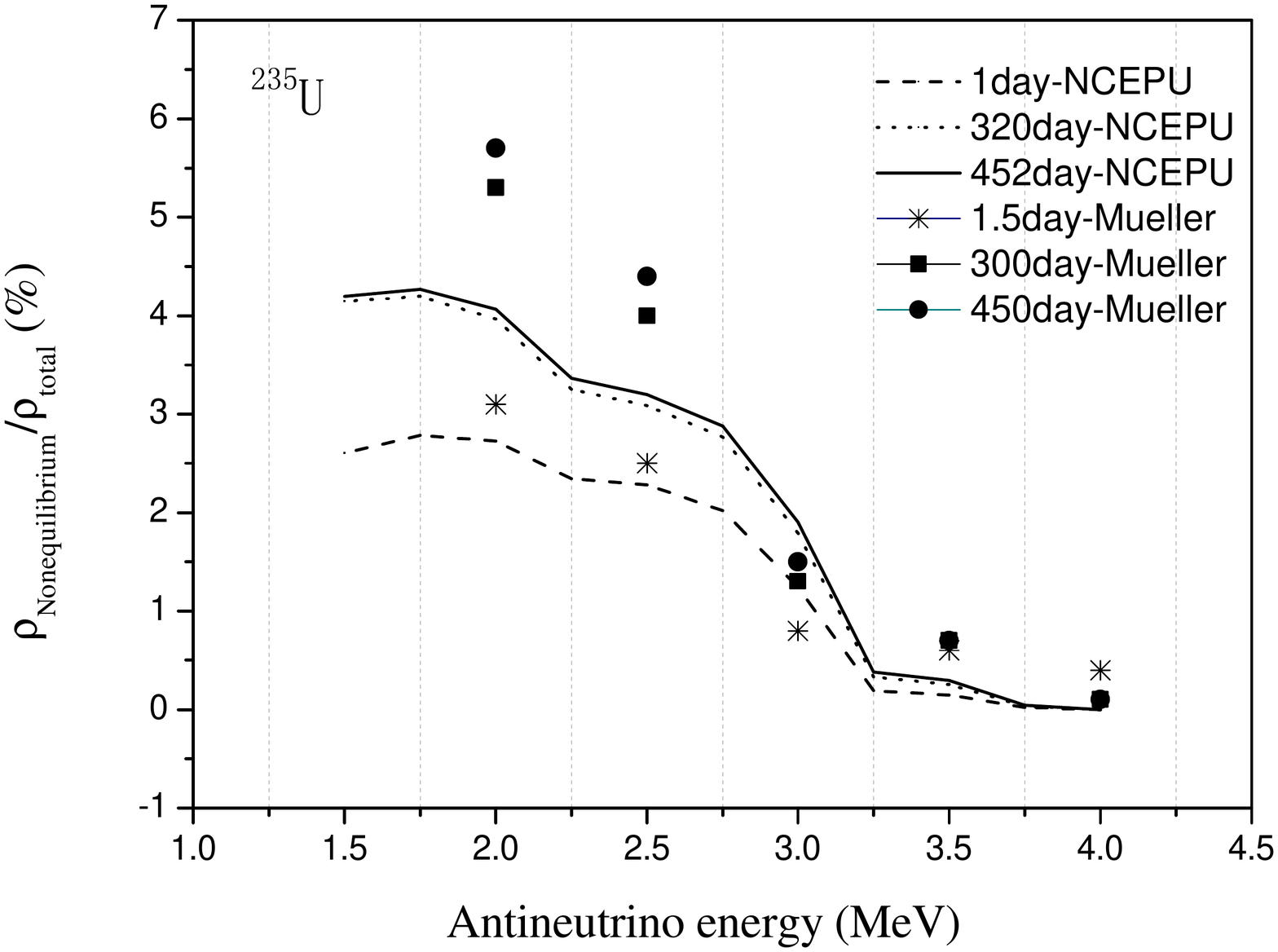}
\caption{Comparison of $^{235}$U non-equilibrium corrections with Mueller's results}
\label{235U_nonequlibrium}
\end{center}
\end{figure}

\begin{figure}
\begin{center}
\includegraphics[width=9cm]{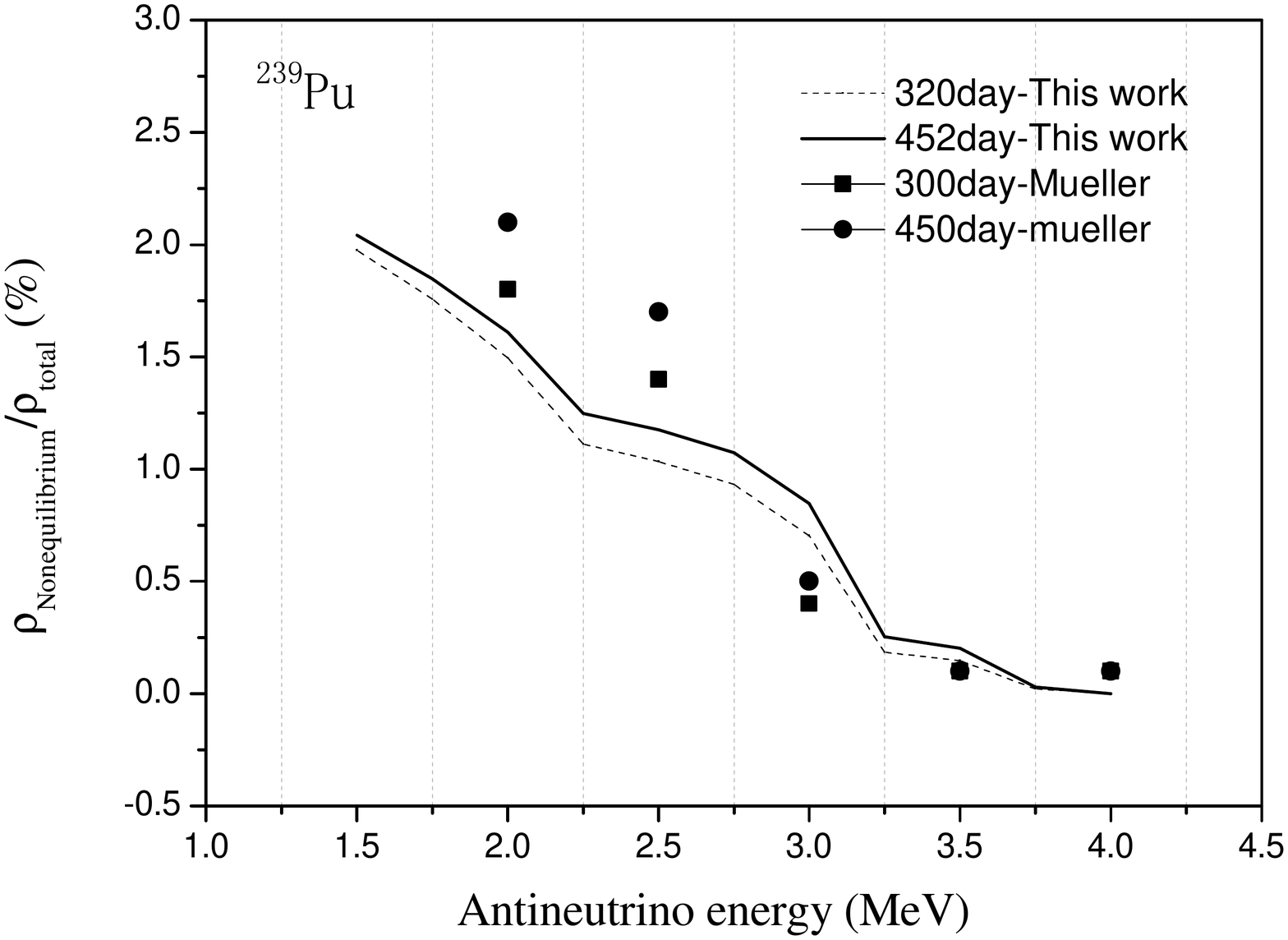}
\caption{Comparison of $^{239}$Pu non-equilibrium corrections with Mueller's results}
\label{Pu239_nonequlibrium}
\end{center}
\end{figure}

\begin{figure}
\begin{center}
\includegraphics[width=9cm]{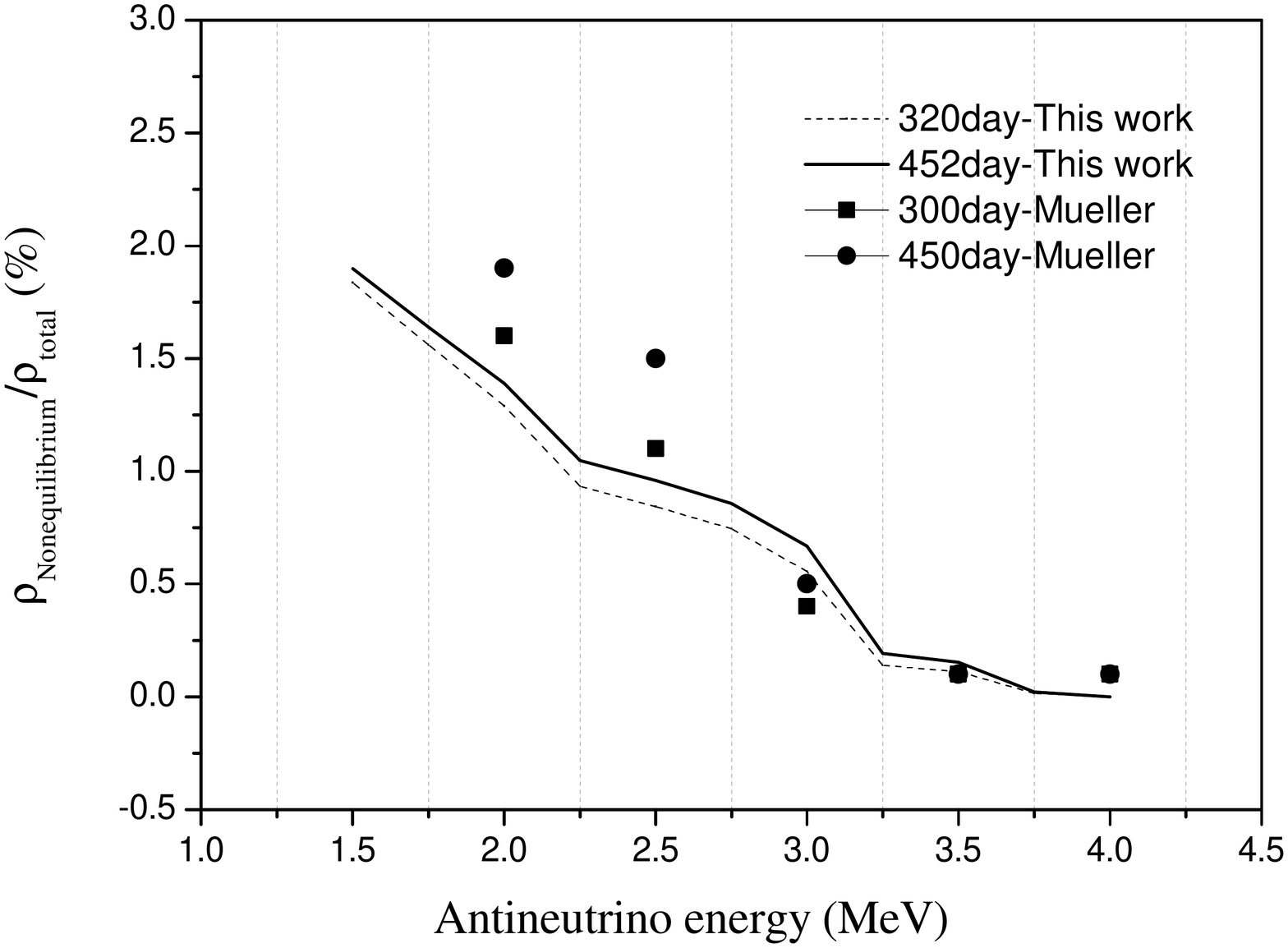}
\caption{Comparison of $^{241}$Pu non-equilibrium corrections with Mueller's results}
\label{Pu241_nonequlibrium}
\end{center}
\end{figure}
\section{Conclusion}
Using all available data regarding SNF, a method to calculate the SNF spectra and non-equilibrium corrections of the ILL spectra was proposed. With more than 2000 isotopes and fission products taken into account in simulations, results show that the SNF contribution to the total antineutrino flux is about 0.26\%$\sim$0.34\%, but the impact from shutdown is about 20\%. The SNF spectrum distorts the softer part of antineutrino spectra, and the maximum SNF contribution is about 1.5\%, but there is an 18\% difference between the results from the line evaluate method and the under evaluate method. Non-equilibrium effects were also calculated using almost the same method as for the SNF calculation. These results are compatible with others considering the uncertainties. The SNF spectrum ratio and the non-equilibrium corrections can be used in future reactor spectra analysis.

\section*{Acknowledgments}
The work was supported by National Natural Science Foundation of China (Grant No. 11390383) and the Fundamental Research Funds for the Central Universities(Grant No. 2015ZZD12), we would like to thank C. Lewis for her verification work, Liang Zhan for his inverse beta decay cross-section calculation code, and Cao Jun for his extraordinary support.






\begin{thebibliography}{10}
%
\bibitem{VKopeikin_2004}V.Kopeikin, L.Mikaelyan, V.Sinev, Antineutrino Background from Spent fuel Storage in sensitive Searches for $\theta_{13}$ at Reactors, Phys. Atom. Nuclei (2006) 69: 185, arXiv:hep-ph/0412044v1 3 Dec 2004
\bibitem{Anfeng_spt}AN Fneg-Peng,TIAN Xin-Chun, ZHAN Liang, et al. Chinese Physics C. 2009, Vol.33 {\bf 9}: 711-716.
\bibitem{zhoubin}ZHOU Bin, RUAN Xi-Chao, NIE Y ang-Bo, et.al. A study of antineutrino spectra from spent nuclear fuel at Daya Bay, CPC(HEP\& NP), 2012, 36(1): 1-5
\bibitem{patricprl}Patrick Jaffke, Corrections to and Applications of the Antineutrino Spectrum Generated by Nuclear Reactors ,Virginia Tech,2015
\bibitem{Anfp_phd} F.P. An, Daya Bay Reactor Neutrino Flux Calculation and Impact on Sensitivity of the Experiment, Institute of High Energy Physics, 2012
\bibitem{VKopeikin_2001}V.I.Kopeikin,L.A.Mikaelyan,V.V.Sinev. Inverse Beta Decay in a Nonequilibrium Antineutrino Flux from a Nuclear reactor, arXiv:hep-ph/0110290v1 23 Oct 2001
\bibitem{CLewis} C.Lewis, Measuring the antineutrino spectrum at the Daya Bay nuclear reactors,University of Wisconsin-Madison,PHD thesis,2014,p79:80
\bibitem{origenarp} I. C. Gauld£¬S. M. Bowman£¬J. E. Horwedel. ORIGEN-ARP: Automatic rapid processing for spent fuel depletion, decay,and source term analysis. USA: ORNL, 2006:13-14.
\bibitem{rmc}Wang Kan, Li Zeguang, She Ding, et.al. Progress on RMC: A Monte Carlo neutron transport code for reactor analysis, International conference on Mathematic and Computational Method(M\&C 2011). Brazil,2011
\bibitem{Endfb7}ENDF/B-VII.0 library, https://www-nds.iaea.org/exfor/endf.htm
\bibitem{arprmc1}G.Ilas, I.C.Gauld, F.C.Difilippo,et al. Analysis of Experimental Data for high Burnup PWR Spent Fuel Isotopic Validation-Calvert Cliffs,Takahama, and Three Mile island reactors. NUREG/CR-6968,ORNL/TM-2008/071, 2009
\bibitem{arprmc2} C. E. Sanders, I. C. Gauld, ORNL, Isotopic Analysis of High-Burnup PWR Spent Fuel Samples From the Takahama-3 Reactor, NUREG/CR-6798,ORNL/TM-2001/259, 2002
\bibitem{muller}Th.A.Mueller, D.Lhuillier, M.Fallot, et al. Improved prediction of reactor antineutrino spectra. Physical Review C,2011, {\bf 83}:054615
\bibitem{lab17}I.C. Gauld, J.C. Ryman, ORNL/TM-2000/284, 2001
\bibitem{hayes}A.C.Hayes, J.L.Friar, G.T.Garvey,et.al. Systematic Uncertainties in the Analysis of the Reactor Neutrino Anomaly, PRL 112,202501 (2014);A.C.Hayes, Petr Vogel, Reactor Neutrino Spectra, 1605.02047v1 [hep-ph] 6 May 2016

\bibitem{patric_huber}Patric Huber, On the determination of antineutrino spectra from nuclear reactor, Physical Review {\bf C85}: 029901,2012
\bibitem{kopei20041} V.I.Kopeikin, L. A. Mikaelyan, and V.V.Sinev, Phys.At.Nucl.67, 11 (2004).
\bibitem{kopei20042}T.R.England and B.F.Rider, LA-UR-94-3106, ENDF-349, LANL (Los Alamos, 1994)


\end{thebibliography}
\end{document}